# KEY PERFORMANCE INDICATORS FOR QOS ASSESSMENT IN TETRA NETWORKS

José Darío Luis Delgado[1] and Jesús Máximo Ramírez Santiago[2]

[1]Department of RF Planning and Optimization, Canal de Comunicaciones Unidas, Madrid, Spain
[2]Department of RF Planning and Optimization, Canal de Comunicaciones Unidas, Madrid, Spain

## ABSTRACT

*Key Performance Indicators (KPIs) are widely used by GSM and UMTS carriers with the aim of evaluating the network performance and the Quality of Service (QoS) delivered to users. TETRA networks are basically designed to provide telecommunication services to Public Safety & Security (PSS) organizations, thus the compliance of the QoS levels required by these clients is usually critical. Despite that, the use of KPIs to assess the network performance and the QoS achieved in TETRA systems is not very common. This paper not only states the need of monitoring and evaluating these parameters, but also introduces a set of KPIs which is considered necessary and sufficient in order to allow TETRA operators to be aware of whether provided services meet the QoS requirements established by end users.*

## 1. INTRODUCTION

As far as telecommunication services for Public Safety & Security (PSS) organizations are concerned, Quality of Service (QoS) requirements are highly demanding and their compliance is critical owing to the type of users of these services. As a result, TErrestrial Trunked RAdio (TETRA) networks are commonly used to provide service to these clients. Despite that, currently there is a lack of publications and proceedings related to the QoS assessment in TETRA networks, unlike for other radio access technologies such as GSM and UMTS. Due to this need, this paper introduces a set of Key Performance Indicators (KPIs) that could be used to assess the QoS levels provided by TETRA networks. The collection of KPIs here proposed constitutes a useful tool for TETRA carriers and planning engineers who require improving the performance of their networks.

The paper is organized as follows: Chapter 2 states an overview of the architecture, features and services provided by TETRA networks. Chapter 3 introduces some concepts related to the QoS assessment in mobile networks, as well as their application to TETRA. Chapter 4 presents a proposal that collects a set of key performance indicators for QoS evaluation in TETRA systems. Finally, Chapter 5 contains the conclusions and the future work.

## 2. TETRA NETWORKS OVERVIEW

### 2.1. TETRA Standard

TETRA is a narrow band mobile communication technology which is specially designed to meet the communication demands from Public Safety & Security (PSS) organizations. Nevertheless, due to their high performance and optimal use of radio resources, TETRA networks are also used in other professional environments such as oil production, electrical companies, terrestrial courier, etc. Actually, TETRA is an open standard developed by the ETSI [1] which specifies this type of cellular radio network, based on trunking channel allocation and TDMA air interface access. The users of TETRA networks are managed into private groups known as organizations or fleets. Obviously, TETRA ensures the absolute privacy in the inter-fleet communications, although there could be some scenarios in which organizations demand combined communications for mutual-aid coordinated services.

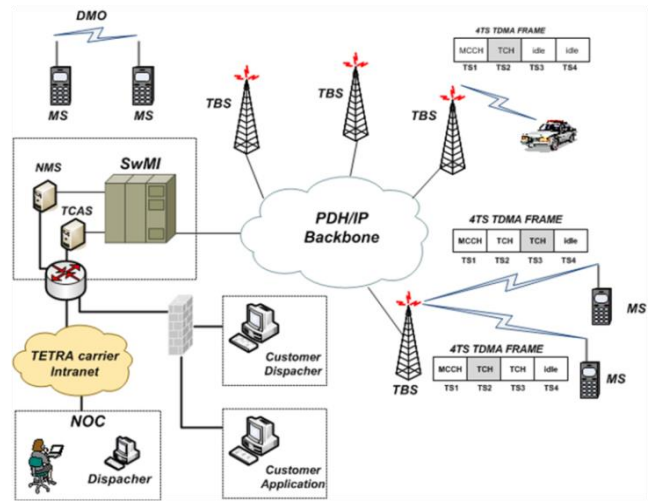

Figure 1. Diagram of the main components of a TETRA network

The generic architecture of a TETRA network is shown in Figure 1. The main components of this type of network technology are outlined below:

- *Switching and Management Infrastructure (SwMI)*: The SwMI is the network core. It is the element responsible for the establishment and switching of communications between users (voice and data), as well as remote management and configuration of base stations, traffic information, billing data generation, performance counters collection, management of access and communication rights of users and organizations, subscriber location register, alarm reporting, etc. All data about users, organizations, configuration of network elements, etc. is stored in the SwMI.

- *TETRA Base Station (TBS)*: The TBS provides the required air interface protocol for subscriber access to the resources and services offered by the system. This protocol is established through the Main Control CHannel (MCCH), a logical channel for signalling data exchange with the Mobile Subscribers (MSs). Every TBS has a MCCH carried on slot 1 of one of its carriers. Depending on the infrastructure supplier, the communication between the TBSs of the network and the SwMI could be carried out by means of either PDH or TCP/IP protocols.

- *Mobile Subscriber (MS)*: The mobile subscriber devices are divided into: hand-held terminals (for personal use), mobile terminals (for installation in vehicles) and fixed terminals (mobile terminals for installation in fixed sites, such as police or fire stations). Each MS has an Individual TETRA Subscriber Identity (ITSI), whereby it is identified in the system. A TETRA user equipment could operate, according to the standard and depending on its configuration, in four different modes:

  o  *TETRA MOde (TMO)*: Communications between MSs are established through the TETRA network infrastructure (TBS+SwMI).

  o  *Direct MOde (DMO)*: Communications between MSs are established directly (end-to-end), without the control of the TETRA infrastructure.

  o  *Gateway Mode*: An equipment configured in this mode operates as a gateway between a group of users working in TMO and another group working in DMO, allowing the information exchange between both modes.

  o  *Repeater Mode*: An equipment configured in this mode operates as a DMO signal repeater to which the rest of MSs must synchronize, allowing the increasing of the coverage distance between DMO users.

- *Network Management Server (NMS)*: This supplementary server is used for system management and remote monitoring of network elements, among other issues.

- *TETRA Customer Application Server (TCAS)*: This server provides the possibility of developing customer applications for system users, such as Automatic Vehicle Localization (AVL), database querying, telemetry, telecontrol, etc.

- *Dispatcher Work Station (DWS)*: These workstations are used to manage and configure users, groups and organizations, as well as to establish communications with subscribers. Depending on the user rights, a DWS could provide a total control (for network operators) or only the control of a few features and options (for organizations).

Despite the similarities between TETRA and other mobile radio networks such as GSM, TETRAPOL, DMR, etc, this technology provides certain features and services which differentiate it and are briefly described in the following sections [2] [3].

## 2.2. Air Interface, Services and Resource Management

According to the standard [1], TETRA air interface is specified as:

- *Frequency bands*: The most widely used frequency bands are 380 - 400 MHz, 410 - 430 MHz and 450 - 470 MHz.

- *Carriers and bandwidth*: FDD carriers with a bandwidth of 25 KHz and duplex spacing of 10 MHz.

- *Access schema and framing*: TDMA with 4 physical time slots (TSs) per carrier. The TETRA frame is composed of these 4 TS.

- *Modulation*: $\pi$/4-DQPSK (linear modulation).

- *Voice codec*: ACELP (4.56 kbps net, 7.2 kbps gross).

- *User Data Rate*: Gross rate of 7.2 kbps per TS and a maximum of 28.8 kbps per carrier (protected up to 19.2 kbps). In TETRA Enhanced Data Service (TEDS) these throughputs could be increased.

- *Logical Channels*: Several logical channels, both for control and traffic information exchange: Main Control CHannel (MCCH), Secondary Control CHannel (SCCH), Associated Control CHannel (ACCH), Fast ACCH (FACCH), Slow ACCH (SACCH), Traffic CHannels (TCHs), Packet Data CHannel (PDCH), etc.

- *Air Interface Encryption (AIE)*: Ciphering of signalling, coded speech and data messages sent on the radio path, by means of encryption keys and algorithms, ensuring the privacy and confidentiality of communications between radio subscribers.

TETRA infrastructures offer specific services and features such as group calls, half-duplex and full-duplex individual calls, broadcast calls, emergency calls, direct mode operation, queued calls, subscriber authentication, Short Data Service (SDS), status messages, customer software applications, management of users and organizations, interconnection with PABX/PSTN and with other TETRA networks, etc. Due to these systems are usually focused on the provision of telecommunication services to PSS users, whose compliance is critical, the performance requirements in TETRA networks are higher than for other type of networks. For this reason, the standard sets some features to improve the management and control of radio resources: call priorities, pre-emptive speech items and calls, priority cells, handover caused by cell load, subscriber classes, configuration of maximum call and speech item duration, limited use of TCHs for data communication, etc.

## 2.3. Registration and Mobility

TETRA networks implement registration and handover processes to provide access to the system and subscriber mobility along the network coverage area, as described below:

### 2.3.1. Network Registration

In order to access the resources and services the network provides, TETRA users have to perform a registration process in which the SwMI usually requires the authentication of MSs. This process is based on the exchange of a key that is provided by the MS manufacturer and must be linked to the associated ITSI in the SwMI, ensuring the wholeness of the subscribers who access the network. A MS is considered as successfully registered into the system when the SwMI has assigned a TBS as its server base station after a switch-on registration or handover process, and the MS is synchronized with the TBS through its MCCH.

### 2.3.2. Group Registration

A MS must be previously registered into a conversational group in order to be able to participate in group calls, which are detailed in the next section. In TETRA networks, groups are identified by means of a Group TETRA Subscriber Identity (GTSI). Generally, MSs are configured including several groups, thus the end user, by selecting the desired group call through the radio device interface, sends a group membership request to the SwMI. If the subscriber is allowed to be a member of the group, according to some previously defined relationships between users and groups in the system, the attachment is accepted, otherwise the request is rejected and the MS could not be involved in group communications. The SwMI is also usually entitled to send group registration demands to MSs through DWSs, allowing the group attachment without the involvement of the user.

### 2.3.3. Handover

The handover process entails that a MS leaves its server TBS and registers into a new TBS which offers it better service quality. The TETRA standard states three events in the radio connection between a server cell and a MS that trigger a handover process: 1) radio link quality degradation, 2) weak signal strength and 3) traffic congestion in the cell. After detecting any of these events, the MS starts a handover process.

A cell handover is considered as successful if the MS achieves to register into a new cell, which is selected from the list of neighbour cells that is broadcasted by the server cell. By means of this handover process, the MS is able to continue using the network resources and processing communications without undergoing any service interruption.

## 2.4. Voice Services

The voice services are considered the main services provided by TETRA infrastructures. As it was previously mentioned, these systems allow assigning several call priorities according to user, group or organization requirements to access the network resources, optimizing the use of those resources.

The voice services are classified into group calls, individual calls, broadcast calls and emergency calls:

### 2.4.1. Group Calls

The group call is the basic TETRA voice service but also the most complex to implement effectively. Group calls are half-duplex point-to-multipoint calls that allow establishing communications between a group of users belonging the same organization. Subscribers who want to be involved in the call must be registered into the target group, as previously stated. When a TETRA user pushes the Push To Talk (PTT) key in its radio device, the group call establishment is carried out without waiting for the acceptance of the receiver subscribers.

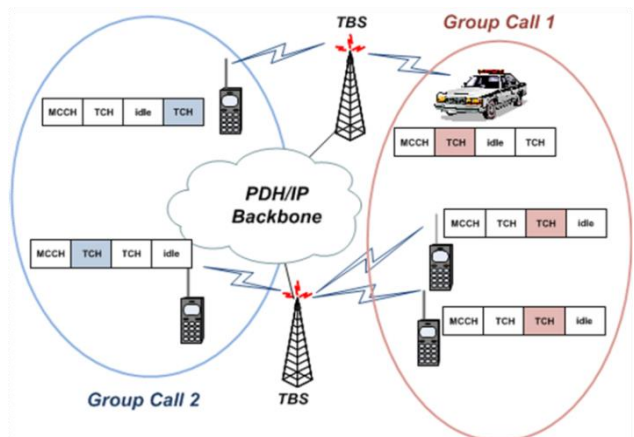

Figure 2. TETRA group call

Group calls are processed by every TBS in which at least one group member is registered into, thus one TCH (one TS) is allocated in each one of these TBSs. If several group members are registered into the same TBS, only one TCH will be allocated as well. Therefore, group calls provide an efficient use of radio resources. In Figure 2 is shown an example of a group call in which two members are registered into the same TBS and process traffic through the same TCH.

### 2.4.2. Individual Calls

Individual calls are point-to-point communications between two subscribers, which could be either half-duplex or full-duplex. To be established, the called party must previously accept the call request from the calling user. The TCHs allocation in the involved TBSs depends on the call type, although if both MSs are registered into different cells, channel allocation is the same regardless of the type: one TCH in each TBS. However, if MSs are registered into the same cell and the call is full-duplex, two TCHs will be allocated, one for each MS, whilst if the call is half-duplex, the allocation is identical to the group call case: only one TCH. Therefore, the use of resources for half-duplex calls is often lower than for full-duplex calls.

### 2.4.3. Broadcast Calls

Broadcast calls are a type of group calls whereby a privileged user is able to establish communication with all users of the organization, ensuring the call reception regardless of whether MSs are processing other voice communications or not. In broadcast calls only the calling party is entitled to transmit, while the rest of subscribers are only allowed to receive the information. Broadcast groups are usually configured in the receiver MSs in such a way that they are not accessible to users.

### 2.4.4. Emergency Calls

Emergency calls are focused on their generation by users who are in real emergency situations, thus these calls have the highest access priority to the network resources, i.e., if the involved TBSs have traffic congestion, the SwMI will interrupt other calls in progress in order to obtain enough resources to process the emergency call. The call type could be dynamically configured in the SwMI as individual or group call, as well as its subscriber or group destination, so that users only need to push the emergency key in the radio device and the call is started without performing any other action.

## 2.5. Data Services

TETRA provides a data exchange service by means of two different procedures: data messages sending (SDS and Status) through the MCCH and data packet sending through the TCHs. A short description of the TETRA data services is presented below:

### 2.5.1. SDS and Status Messages

TETRA standard defines two types of data messages: the Short Data Service (SDS) messages and the Status messages, which are sent through the MCCH. The former are data messages which could be exchanged between MSs, DWSs or customer software applications, in either point-to-point or point-to-multipoint communication. The standard states four types of SDS messages depending on their bit length, from SDS Type 1 up to SDS Type 4. SDS messages could also be used as data transport mechanism for different customer applications, such as AVL. Status messages are 16 bits-long messages, thus their sending requires a minimum use of system resources. Up to 65535 numerical values could be encoded, being a range excluded for user applications because it is exclusively reserved for network use.

### 2.5.2. Packet Data Mode

Besides the use of TCHs for voice communications, they could be used for IP data packets transfers using a specific channel known as Packet Data CHannel (PDCH). Hence, the SwMI provides the possibility to establish TCP/IP connectivity between subscribers and customer applications or with others networks. Nevertheless, the packet data service is not normally used because it offers a low transmission throughput and makes use of radio resources that are required for other critical services.

### 2.5.3. Customer Applications

TETRA systems provide a set of APIs (DLL libraries) through the TCAS, which allow the development of customer software applications (AVL, telemetry, telecontrol, etc.).

## 3. QOS ASSESSMENT IN WIRELESS NETWORKS

ITU-T Rec. E.800 [4] and Rec. G.1000 [5] collect a set of definitions related to the quality of a telecommunication service. Definitions of service, quality of service and quality of network performance provided by these guidelines are shown below:

- *Service*: Set of functions that a telecommunication system provides to the user.

- *Quality of service (QoS)*: Totality of characteristics of a telecommunication service that bear on its ability to satisfy stated and implied needs of the user of the service.

- *Quality of network performance*: Network ability to offer the functions matching the communications between users. The quality of network performance contributes towards the QoS as experienced by the user. In mobile communication systems, network performance may be evaluated according to an end to end basis or not, in which scenario the quality of the transmission network and access network must be separated.

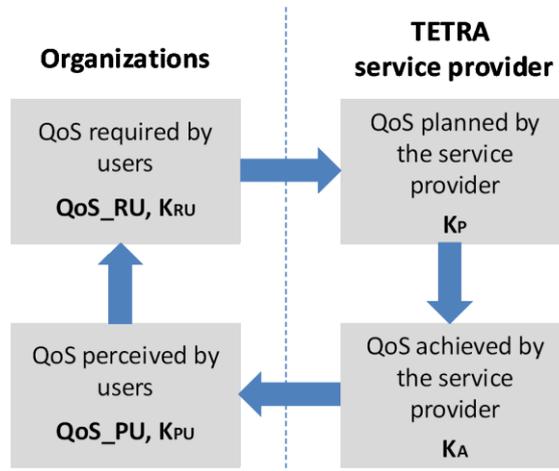

Figure 3. The four perspectives of the QoS

The quality of service could be broken down into four viewpoints or perspectives, as shown in Figure 3:

1. *QoS required by the customer*: It states the quality levels the user requires from a particular telecommunication service. These requirements could be expressed in non-technical language by means of descriptive terms and criteria that the service provider must translate into QoS parameters fitting the service. Hence, for TETRA networks is necessary to find a relation $f(\cdot)$ to translate the $L$ requirements for each one of the $N$ organizations in the network, denoted as $QoS\_RU \in \mathbb{R}^{N \times L}$ (QoS Required by Users), into $M$ quantifiable QoS parameters, denoted as $k_{RU} \in \mathbb{R}^{M}$, in such a way that:

$$k_{RU} = f(QoS\_RU).$$

$QoS\_RU$ is the matrix of QoS requirements from clients, defined as:

$$QoS\_RU = \begin{bmatrix} qos\_ru_{1,1} & \cdots & qos\_ru_{1,L} \\ \vdots & \ddots & \vdots \\ qos\_ru_{N,1} & \cdots & qos\_ru_{N,L} \end{bmatrix},$$

where $qos\_ru_{i,j}$ is the $j-th$ QoS requirement from the $i-th$ organization. Finding $f(\cdot)$ is not a straightforward task and it depends on the type of services demanded by users, as well as on the possible differences between organizations regarding the assignment of service priorities, among other issues.

2. *QoS planned by the service provider*: It is a statement of the level of quality that the service provider expects to offer to the customer. It is expressed by means of values that are assigned to specific QoS parameters. Ideally, the service provider should define the QoS planned, denoted by means of the $k_P \epsilon \mathbb{R}^M$ vector, to be equal to $k_{RU}$.

3. *QoS achieved or delivered by the service provider*: It is a statement of the level of quality actually achieved and delivered to the customer. It is expressed by means of values that are assigned to some specific parameters of QoS, which should be the same as specified for the planned QoS. The QoS achieved, denoted by means of the $k_A \epsilon \mathbb{R}^M$ vector, is obtained through some performance indicators gathered from the network, which constitute the principal subject of this work.

4. *QoS perceived by the customer*: It is a statement whereby the users express the service quality levels they believe have experienced, which are usually indicated in terms of satisfaction degrees and not in technical terms. The QoS levels perceived by users should be translated into QoS parameters to fit the rest of perspectives. Hence, for TETRA networks is necessary to find a relation $g(\cdot)$ to translate the QoS levels perceived by users, denoted as $QoS\_PU \epsilon \mathbb{R}^{NxL}$ (QoS Perceived by Users), into $M$ quantifiable QoS parameters, denoted as $k_{PU} \epsilon \mathbb{R}^M$, in such a way that:

$$k_{PU} = g(QoS\_PU)$$

QoS_PU is the matrix of the QoS perceived by clients, defined as:

$$QoS\_PU = \begin{bmatrix} qos\_pu_{1,1} & \cdots & qos\_pu_{1,L} \\ \vdots & \ddots & \vdots \\ qos\_pu_{N,1} & \cdots & qos\_pu_{N,L} \end{bmatrix},$$

where $qos\_pu_{i,j}$ is the $j-th$ QoS perception from the $i-th$ organization. For simplicity, $f(\cdot)$ and $g(\cdot)$ could be considered the same, so that:

$$k_{PU} = f(QoS\_PU).$$

One of the most important issues that must be considered as starting point to plan and design a wireless access network, consists of the evaluation of QoS needs and requirements from future customers. This set of requirements must contain all information needed by the service provider in order to determine the QoS that should be planned and offered to network users. The combination of the relationships between the four perspectives of the QoS sets up the basis of an effective and practical QoS management, thus it could be stated that QoS is getting better when the four perspectives for a specific service are reaching the convergence. Hence the need for the service provider to have some tools to analyze the network performance and allow the assessment of the quality levels provided to users. In those cases in which actual quality levels divert from the theoretical planned values, the service provider must carry out some tasks with the aim of maximizing the provided quality levels, to achieve the convergence between the four perspectives of the QoS, and optimizing the use of available radio resources.

Due to all above-mentioned, when the network is deployed and the services are being provided to radio subscribers, it is necessary to introduce some systems to monitor and assess the QoS levels delivered to the end user. In TETRA networks, these systems use some counters of events that are collected and recorded by the SwMI. These counters are commonly known as performance counters and gather most events that occur in the access network during an observation time that must be configured by network carriers. When recorded, the counters must be processed to obtain some indicators known as Key Performance Indicators (KPIs), which provide more useful information to analyze the achieved QoS and network performance. Based on the estimation of these indicators, it must be generated a set of periodic reports of QoS according to the desired temporal scale: daily, weekly, monthly, etc., which will allow the service provider to get an overall assessment of the network status and provided services. The relationship between event counters, KPIs and QoS reports is shown in Figure 4.

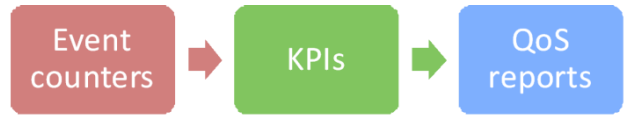

Figure 4. Event counters, KPIs and QoS reports

By means of the monitoring of key performance indicators and the assessment of QoS reports, the service provider could achieve, among others, the following targets:

1. Identifying occasional faults in the hardware equipment of base stations.

2. Detecting some interference problems and service degradations, that would motivate the carrying out of corrective actions such as new frequency allocations, adjustment of antennas or amendment of radio parameters.

3. Locating some congestion or over-dimensioning of the capacity problems, both for MCCHs and TCHs, which would entail the performance of a more suitable dimensioning according to what users demand.

4. Monitoring the network performance, as well as its variations and trends, allowing the service provider to plan some preventive actions in that regard.

On the other hand, performing drive test measurements allows the assessment of the network coverage and radio link quality provided by base stations (signal strengths, C/I ratios or error rates, among other parameters). Altogether, from the drive test measurements and the QoS reports, it is possible to have an overall and sufficient view of the network status and the service quality levels delivered to users. Thereby, the service provider could correct the possible QoS deviations regarding the designed values, with the ultimate aim of reaching the convergence between the four perspectives of the QoS.

## 4. PROPOSED KEY PERFORMANCE INDICATORS FOR TETRA NETWORKS

Despite the fact that monitoring and evaluating the key performance indicators seems basic within the processes of design and optimization of any kind of network, in the case of TETRA systems, and unlike for GSM networks [6] [7], there are hardly any researches and publications so far which collect some metrics in order to assess the service quality levels provided to users. As TETRA is a radio communication network, most parameters to quantify in these systems is similar to those applied to other radio access technologies designed primarily for the provision of voice services. However, due to their own features, it is necessary to make a specific proposal for TETRA systems. Some generic indicators for mobile networks are proposed in [8], where those applicable to TETRA networks are remarked. Nevertheless, most indicators offered by this guideline are oriented to their implementation under the viewpoint of infrastructure manufacturers, and not to their estimation by network carriers.

Throughout this chapter the main performance counters that TETRA infrastructures usually provide are described, from which it is proposed a collection of KPIs that could be considered necessary and sufficient to decide whether offered services meet the minimal requirements that users demand. These indicators are introduced gathered in several categories according to the parameters the network operator desires to measure and assess.

## 4.1. Availability Indicators

They are defined similarly to the availability indicators for other mobile networks, i.e., they give information about the time percentage during which a base station is providing service. The availability indicator for a TBS could be directly obtained through a counter that collects the total time the base station is providing service during the observation time. From this data, some availability reports of base stations could be generated according to the desired temporary scale.

## 4.2. Network Resource Indicators

These indicators provide information related to the degree of use of the network radio resources, i.e., traffic and control channels, thus they also could be denominated as accessibility indicators. TETRA networks are systems of queued calls, in which the traffic capacity is modelled by applying the Erlang C distribution [9]. In this type of systems, when a channel reservation is requested for voice or data services and the base station does not have enough resources to allocate, the request is queued until some resources are released or the maximum queuing time the service provider has configured is expired.

These KPIs allow the assessment of the dimensioning of the base stations capacity, with the aim of identifying areas in which traffic demands could make necessary an enlargement of the initially planned capacity, or maybe could be considered advisable to downsize such capacity. To analyze the congestion of a base station, it must be determined the daily hour in which it has been processed the maximum volume of voice and data packets traffic, which is known as Busy Hour (BH), and it must be evaluated the number of queued requests that occurred at that hour. The assessment of the congestion at BH avoid analyzing occasional peaks generated by sporadic occurrences of several simultaneous call attempts, which do not match the daily hour of maximum traffic processed by the base station.

The most common event counters related to the use of network resources are collected below. These counters are usually recorded for every base station in TETRA networks during an observation time configured by the operator. In parenthesis is shown an acronym associated with each counter, which will be used later within the proposed mathematical expressions for KPI estimation:

- *Group calls busy time (GCBT)*: Busy time of traffic channels due to the process of group calls.

- *Individual calls busy time (ICBT)*: Busy time of traffic channels due to the process of individual calls.

- *Packet mode busy time (PMBT)*: Busy time of traffic channels due to the transfer of data packets.

- *Provided capacity (PC)*: Total time the base station provides in order to process voice calls and packet mode data.

- *Channel reservation requests (CRR)*: Number of channel reservation requests performed by MSs to establish voice calls and packet mode transfers.

- *Queued channel reservation requests (QCRR)*: Number of channel reservation requests which are placed in the waiting queue.

- *Queuing waiting time (QWT)*: Total waiting time for all queued voice calls and data packets.

- *Peak of simultaneously queued requests (PSQR)*: Maximum number of channel requests which are waiting simultaneously in the queue.

- *Downlink MCCH sent semislots (DSS)*: Number of available semislots in the downlink control channel to send signalling, system broadcasts and SDS and status messages.

- *Downlink MCCH used semislots (DUS)*: Number of used semislots in the downlink control channel to send signalling, system broadcasts and SDS and status messages.

- *Uplink MCCH reserved semislots (URS)*: Number of semislots reserved by the base station in the uplink control channel to send signalling and SDS and status messages.

- *Uplink MCCH used semislots (UUS)*: Number of used semislots in the uplink control channel to send signalling and SDS and status messages.

- *Random access collisions (RAC)*: Number of semislots in the uplink control channel in which collisions are detected due to the simultaneous random access from several MSs. It denotes the existence of overload of data messages or interferer signals with non-TETRA modulation.

The proposed network resource indicators are defined next, as well as the mathematical expressions used for their estimation, which are obtained from the previous event counters that are usually provided by TETRA networks:

- **Voice occupancy at BH (%)**:

  It denotes the degree of occupancy of the available traffic channels in the base station at BH, due to the process of voice communications, i.e., group calls and individual calls.

  $$Voice\ Occup\ BH\ (\%) = \frac{GCBT + ICBT}{PC}\ x\ 100.$$

- **Data occupancy at BH (%)**:

  It denotes the degree of occupancy of the available traffic channels in the base station at BH, due to the process of packet mode data communications.

  $$Data\ Occup\ BH\ (\%) = \frac{PMBT}{PC}\ x\ 100.$$

- **Total traffic occupancy at BH (%)**:

  It denotes the degree of occupancy of the available traffic channels in the base station at BH, due to the process of voice and packet mode data communications.

  $$Occup\ BH\ (\%) = \frac{GCBT + ICBT + PMBT}{PC}\ x\ 100.$$

- **Efficiency of use of the network capacity (%)**:

  It shows the use of the traffic channels by means of a comparative between the traffic processed by the network and the capacity it offers, and allow evaluating if the provided network capacity fits the actual needs from users. It is recommended to divide the network into different clusters and to include in every cluster several base stations, thus the use of the network capacity for every cluster $C_i$ could be evaluated by means of the following equation:

  $$K_i(\%) = \frac{Real\ Traffic_i}{Theoric\ Traffic_i}\ x\ 100,$$

  where $Theoric\ Traffic_i$ is the addition of the traffic capacity of every TBS inside the $i-th$ cluster regarding one hour, and $Real\ Traffic_i$ is the addition of the average of traffic occupancy at BH during the measurement period for every TBS inside the $i - th$ cluster.

- **Queuing rate at BH (%)**:

  It provides the ratio between the number of voice calls and data packets that are placed in the waiting queue at BH, and the total number of traffic channel reservation requests that occur at that hour.

  $$Queuing\ Rate\ BH\ (\%) = \frac{QCRR}{CRR}\ x\ 100.$$

- **Mean queuing time at BH (s)**:

  It provides the mean time a voice call or data packet transfer request must wait in the queue until getting the allocation of a traffic channel to establish communication.

  $$Mean\ Queuing\ Time\ BH\ (s) = \frac{QWT}{QCRR}.$$

- **Peak of simultaneously queued requests at BH**:

  It shows the maximum number of channel requests that have to wait in the queue simultaneously at BH.

  $$Simult\ Queued\ Requests\ Peak\ BH = PSQR.$$

- **Daily maximum occupancy of the downlink MCCH (%)**:

  It denotes the daily maximum level of occupancy of the downlink main control channel, due to the sending of signalling, systems broadcasts and SDS and status messages. It must be estimated for the values of the counters matching the daily hour in which DUS achieves its maximum value.

  $$Max\ Occup\ MCCH\ DL\ (\%) = \frac{DUS}{DSS}\ x\ 100.$$

- **Daily maximum occupancy of the uplink MCCH (%)**:

  It denotes the daily maximum level of occupancy of the uplink main control channel, due to the sending of signalling and SDS and Status messages. It must be estimated for the values of the counters matching the daily hour in which UUS achieves its maximum value.

  $$Max\ Occup\ MCCH\ UL\ (\%) = \frac{UUS}{URS}\ x\ 100.$$

- **Random access collisions on the MCCH**:

  It shows the number of semislots in the uplink main control channel in which collisions have been detected due to the simultaneous random access. The associated counter could also increase due to the detection of signals with non-TETRA modulation within the frequency band of interest. It is recommended to consider the counters which have values above a specific threshold fixed by the network carrier.

  $$Random\ Access\ Collisions\ MCCH = RAC.$$

## 4.3. Group Attachment Indicators

This group of indicators provide information related to the degree of success of the group attachment process, considering both the requests initiated by MSs and those that are carried out on system demand. An unsuccessful group registration request could be caused by some problems such as weak quality of the radio link or failures in network elements.

The most common event counters regarding the group attachment process are collected below. These counters are usually recorded for every base station in TETRA networks during an observation time configured by the operator:

- *Group attachment requests initiated by MSs (GAU)*: Total number of group attachment requests carried out on user demand, both successful and unsuccessful.

- *Unsuccessful group attachment requests initiated by MSs (UGAU)*: Number of unsuccessful group attachment requests carried out on user demand.

- *Group attachment requests initiated by the SwMI (GAS)*: Total number of group attachment requests carried out on system demand, both successful and unsuccessful.

- *Unsuccessful group attachment requests initiated by the SwMI (UGAS)*: Number of unsuccessful group attachment requests carried out on system demand.

The proposed group attachment indicators are defined following, as well as the mathematical expressions used for their estimation, which are obtained from the previous event counters that are usually provided by TETRA networks. It is recommended to consider the daily worst case for every indicator, i.e., the case matching the higher failure rate, and base station.

- **Failure rate in group attachments required by MSs (%)**:

  It provides the ratio between the number of unsuccessful group attachments and the total number of group attachment requests carried out by mobile subscribers.

$$MS\ Group\ Attach\ Failure\ Rate\ (\%) = \frac{UGAU}{GAU}\ x\ 100.$$

- **Failure rate in group attachments required by the SwMI (%)**:

  It provides the ratio between the number of unsuccessful group attachments and the total number of group attachment requests carried out by the system.

$$SwMI\ Group\ Attach\ Failure\ Rate\ (\%) = \frac{UGAS}{GAS}\ x\ 100.$$

## 4.4. Handover Indicators

These indicators provide information about the system ability to keep communications between users uninterrupted when they are moving and changing the server cell, thus they allow the assessment of the radio quality and the sustainability of the service during handover processes. Failures that happens during these processes could be caused by problems such as congestion in the target cell or existence of interferences in the network.

The most common event counters regarding the handover process are collected below. These counters are usually recorded for every base station in TETRA networks during an observation time configured by the operator:

- *Individual call handovers (ICH)*: Total number of handover requests carried out by MSs during the processing of individual calls, in order to attach the base station as the new server cell.

- *Unsuccessful individual call handovers (UICH)*: Number of failed handover requests carried out by MSs during the processing of individual calls, in order to attach the base station as the new server cell.

- *Group call handovers (GCH)*: Total number of handover requests carried out by MSs during the processing of group calls, in order to attach the base station as the new server cell.

- *Unsuccessful group call handovers (UGCH)*: Number of failed handover requests carried out by MSs during the processing of group calls, in order to attach the base station as the new server cell.

The proposed handover indicators are defined next, as well as the mathematical expressions used for their estimation, which are obtained from the previous event counters that are usually provided by TETRA networks. It is recommended to consider the daily worst case for every indicator, i.e., the case matching the higher failure rate, and base station:

- **Failure rate in handovers during individual call (%)**:

  It denotes the ratio between the number of failures occurred during handovers while processing individual calls, and the total number of handovers carried out during individual calls with the aim of attaching the base station as the new server cell.

  $$Indiv\ Call\ Handover\ Failure\ Rate\ (\%) = \frac{UICH}{ICH}\ x\ 100.$$

- **Failure rate in handovers during group call (%)**:

  It denotes the ratio between the number of failures occurred during handovers while processing group calls, and the total number of handovers carried out during group calls with the aim of attaching the base station as the new server cell.

  $$Group\ Call\ Handover\ Failure\ Rate\ (\%) = \frac{UGCH}{GCH}\ x\ 100.$$

## 4.5. Voice Service Indicators

This group of indicators provide information related to the network ability to set up calls successfully, as well as to keep them in process without unexpected outages until users hang them up, i.e., these indicators allow the assessment of the radio quality and the sustainability of the service regarding voice calls. Failures in the establishment and processing of calls could be caused by problems such as presence of interferences, faults in network elements or coverage lack of one of the call-involved parties. Moreover, some calls could be released prior to their ending due to the process of pre-emptive priority calls.

The most common event counters regarding the voice service are collected below. These counters are usually recorded for every base station in TETRA networks during an observation time configurable by the operator:

- *Successful placed group calls (SPGC)*: Number of group calls that have been successfully established.

- *Unsuccessful placed group calls (UPGC)*: Number of group calls in which a failure in the set up process happens.

- *Successful placed individual calls (SPIC)*: Number of individual calls that have been successfully established.

- *Unsuccessful placed individual calls (UPIC)*: Number of individual calls in which a failure in the set up process happens. Successful ended group calls (SEGC): Number of group calls that have been successfully ended.

- *Successful ended group calls (SEGC)*: Number of group calls that have been successfully ended.

- *Unsuccessful ended group calls (UEGC)*: Number of group calls that are released before a user hangs up.

- *Successful ended individual calls (SEIC)*: Number of individual calls that have been successfully ended.

- *Unsuccessful ended individual calls (UEIC)*: Number of individual calls that are released before a user hangs up.

The proposed voice service indicators are defined following, as well as the mathematical expressions used for their estimation, which are obtained from the previous event counters that are usually provided by TETRA networks. It is recommended to consider the daily worst case for every indicator, i.e., the case matching the higher failure rate, and base station:

- **Failure rate in group call set up (%)**:

  It provides the ratio between the number of failed group call establishments and the total number of set up requests of this kind of calls processed by the TBS.

  $$Group\ Call\ Setup\ Failure\ Rate\ (\%) = \frac{UPGC}{SPGC + UPGC}\ x\ 100.$$

- **Failure rate in individual call set up (%)**:

  It provides the ratio between the number of failed individual call establishments and the total number of set up requests of this kind of calls processed by the TBS.

  $$Indiv\ Call\ Setup\ Failure\ Rate\ (\%) = \frac{UPIC}{SPIC + UPIC}\ x\ 100.$$

- **Failure rate in group call processing (%)**:

  It provides the ratio between the number of placed group calls that afterwards undergo an unexpected ending, and the total number of group calls whose ending is processed by the TBS.

  $$Group\ Call\ Process\ Failure\ Rate\ (\%) = \frac{UEGC}{SEGC + UEGC}\ x\ 100.$$

- **Failure rate in individual call processing (%)**:

  It provides the ratio between the number of placed individual calls that afterwards undergo an unexpected ending, and the total number of individual calls whose ending is processed by the TBS.

  $$Indiv\ Call\ Process\ Failure\ Rate\ (\%) = \frac{UEIC}{SEIC + UEIC}\ x\ 100.$$

### 4.6. Data Service Indicators

These indicators provide information about the degree of success in the delivery of SDS messages, status messages and data packets through the network infrastructure, considering both those that are sent from dispatcher stations and client applications, and those sent from user devices. Failures in the delivery of messages and errors that happen during the transmission of packets usually denote a weak quality of the radio link.

The most common event counters regarding the data service, both in circuit mode and packet mode, are collected below. These counters are usually recorded for every base station in TETRA networks during an observation time configurable by the operator:

- *Messages sent by DWSs and customer applications (DAM)*: Total number of SDS and status messages which are sent from dispatcher stations and client applications.

- *Non-delivered messages sent by DWSs and customer applications (UDAM)*: Number of SDS and status messages which are sent from dispatcher stations and client applications and are not delivered to their destination.

- *Messages sent by MSs (UM)*: Total number of SDS and status messages which are sent from user equipment.

- *Non-delivered messages sent by MSs (UUM)*: Number of SDS and status messages which are sent from user equipment and are not delivered to their destination.

- *Successful received packets (SRP)*: Number of data packets that are received without errors by the base station.

- *Corrupt received packets (CRP)*: Number of data packets with errors that are received by the base station.

The proposed data service indicators are defined next, as well as the mathematical expressions used for their estimation, which are obtained from the previous event counters that are usually provided by TETRA networks. It is recommended to consider the daily worst case for every indicator, i.e., the case matching the higher failure rate, and base station:

- **Failure rate in the delivery of messages sent from DWSs and customer applications (%)**:

  It shows the percentage of messages sent from dispatcher stations and client applications which have not been successful delivered to their destination.

$$DWS/App\ Sent\ Failure\ Rate\ (\%) = \frac{UDAM}{DAM}\ x\ 100.$$

- **Failure rate in the delivery of messages sent from MSs (%)**:

  It shows the percentage of messages sent from user equipment which have not been successfully delivered to their destination.

$$MS\ Sent\ Faiure\ Rate\ (\%) = \frac{UUM}{UM}\ x\ 100.$$

- **Failure rate in the transmission of data packets (%)**:

  It shows the percentage of packets received with errors by the base station.

$$Packets\ Failure\ Rate\ (\%) = \frac{CRP}{SRP + CRP}\ x\ 100.$$

## 5. CONCLUSIONS AND FUTURE WORK

The collection of KPIs proposed in this paper provides a sufficient metric to assess the performance of a mobile radio network based on TETRA standard, as well as the quality levels with which the services demanded by system users are provided. These indicators make up a basic tool for network operators in order to achieve the convergence between the four perspectives of QoS, meeting thereby the needs and requirements of customers, which are established as design targets during the system planning stage.

Nevertheless, the process of monitoring the indicators does not lead to the knowledge of the compliance degree of the planned quality levels whether it is not followed by a process of assessment of these indicators. Therefore, it must be established a set of target values for all proposed KPIs so that the validity of the provided quality levels could be evaluated during the analysis of QoS reports. However, due to this type of metrics are not widely used in TETRA networks, it is not easy to find some reference values and conditions which could be used as suitable decision thresholds. To overcome this lack of information, in a coming publication it will be shown an assessment proposal for the KPIs presented in this work, achieved from analyzing a real medium size TETRA network which provides service to several organizations over a wide geographic area. Hence, that second work will constitute a reliable guideline that could be taken as a reference by TETRA network carriers in order to help them to assess the system performance and the QoS delivered to end users, concluding thereby the study initiated in this paper.

## REFERENCES

[1] Terrestrial Trunked Radio (TETRA); Voice plus Data (V+D); Part 2: Air Interface (AI), ETSI EN 300 392-2, 2001.

[2] J. Dunlop, D. Girma and J. Irvine, Digital Mobile Communications and the TETRA System. New York, USA: Wiley, 1999.

[3] P. Stavroulakis, Terrestrial Trunked Radio - TETRA. A Global Security Tool. Crete, Greece: Springer, 2007.

[4] Definitions of terms related to quality of service, ITU-T Rec. E.800, 2008.

[5] Communications quality of service: A framework and definitions, ITU-T Rec. G.1000, 2011.

[6] B. Haider, M. Zafrullah and M. K. Islam, "Radio Frequency Optimization & QoS Evaluation in Operational GSM Network," in Proc. World Congr. Engineering and Computer Science, San Francisco, USA, 2009.

[7] Gómez and R. Sánchez, End-to-End Quality of Service over Cellular Networks. Data Services Performance and Optimization in 2G/3G. Chichester, England: Wiley, 2005.

[8] Speech and multimedia Transmission Quality (STQ); QoS aspects for popular services in mobile networks; Part 2: Definition of Quality of Service parameters and their computation, ETSI TS 102 250-2, 2011.

[9] Teletraffic Engineering, ITU-D Study Group 2 Question 16/2, Geneva, Switzerland, 2006.

**Authors**

José Darío Luis Delgado was born in Maracay, Venezuela, in 1971. He received the Electronic Engineering degree (IUPFAN, Maracay) in 1992 and the MSc degree in Electronic Engineering (Universidad Simón Bolivar, Venezuela) with a grant research for his thesis work in 1998. He has been working on several projects related with wireless network telecommunications since 1993, obtaining wide experience on microwave radio-links, PMR, WiMAX, MPT1327 and TETRA networks roll-out and optimization projects. He has also worked for several universities as associated professor as well as researcher engineer. He is currently working as Project Manager for a TETRA network carrier in Madrid. 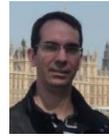

Jesús Máximo Ramírez Santiago was born in Porcuna (Jaén), Spain, in 1987. He received the Telecommunication Engineering degree from the University of Sevilla, Spain, in 2011, specializing in the field of signal theory and wireless communications. He has worked on several projects related to mobile communications for the Technical University of Łódź, Poland, as an exchange grant student. Currently, he is working for a Spanish telecommunication operator on projects of radio planning, design and optimization of TETRA and DMR access networks, among other issues. He is also interested in the field of the Software Engineering, developing mobile applications as a freelance professional. 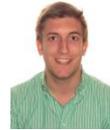